%% file: cairnsproc.tex
\documentclass[fleqn,twoside]{article}
\usepackage{espcrc2}

\usepackage{pictex}
\usepackage[dvips]{graphicx}

\include{commands}

\title{
\vspace{-1.8cm}
\hfill \rm \null \hfill
 \hbox{\normalsize ADP-01-47/T479} \\
\vspace{+1.3cm}
Improving the low-lying spectrum of the overlap kernel}

\author{W. Kamleh\address[CSSM]{Special Research Centre for the Subatomic Structure of Matter (CSSM) and Department of Physics and Mathematical Physics, University of Adelaide 5005, Australia.}\thanks{\tt wkamleh@physics.adelaide.edu.au},
D. Adams\addressmark[CSSM]\thanks{\tt dadams@physics.adelaide.edu.au},
D.B. Leinweber\addressmark[CSSM]\thanks{\tt dleinweb@physics.adelaide.edu.au},
A.G. Williams\addressmark[CSSM]\thanks{\tt awilliam@physics.adelaide.edu.au} }

\begin{document}

\thispagestyle{empty}

\begin{abstract}
The action of the overlap-Dirac operator on a vector is typically implemented indirectly through a multi-shift conjugate gradient solver. The compute-time this takes to evaluate depends upon the condition number $\kappa$ of the matrix that is used as the overlap kernel. We examine the low-lying spectra of various candidate kernels in an effort to optimise $\kappa$, thereby speeding up the overlap evaluation.
\end{abstract}

\maketitle

\section{Introduction}

Overlap fermions are a realisation of chiral symmetry on the lattice. Given some reasonable Hermitian-Dirac operator $H$, we can deform $H$ into a chiral action through the overlap formalism,
\eqn{D_o = \frac{1}{2}\big( 1+\gamma_5 \epsilon(H) \big), \quad \epsilon(H)=\frac{H}{\sqrt{H^2}.} }
Unfortunately, the matrix sign function $\epsilon(H)$ is difficult to evaluate and is typically approximated by a sum over poles \cite{neuberger-practical} which can be evaluated using a multi-shift conjugate gradient (CG) solver \cite{edwards-practical}. This is an iterative approximation where the number of iterations for a given accuracy depends upon the condition number of the kernel, $\kappa(H) = |\lambda_{\rm max}/\lambda_{\rm min}|$.

Usually the Hermitian Wilson-Dirac operator is used as the overlap kernel. Its low-lying spectrum is characterised by a handful of isolated eigenmodes which can be very small, increasing $\kappa$ unacceptably. These eigenmodes can be projected out of the basic operator, reducing its condition number to a numerically acceptable level, and then dealt with explicitly\cite{edwards-chiral}. Unfortunately, as the spectrum rapidly becomes dense, projecting out low-lying modes can only help one so far.

An alternative is to use a kernel with an improved spectrum, that is, where the region of dense modes is shifted away from zero. In this paper we consider six different actions as potential kernels. The first three actions are the standard Wilson with and without the addition of the clover term at both tree-level and with a mean-field improved coefficient. The last three actions are derived from the first three actions by APE-smearing the irrelevant terms. The details of the actions are given in \cite{kamleh-accelerated}, in particular, the FLIC (Fat Link Irrelevant Clover)\cite{zanotti-hadron} action is defined there.

\section{Spectral Flow Comparison}

In order to test the merits of each of our proposed actions, we calculate the spectral flow of each of them to see if our reasoning regarding their low-lying spectra is correct. From the quadratic form of the lower bounds as a function of $m$, and based upon results given in \cite{edwards-spectral}, we expect there to be some peak value of $m$ for which the gap around zero is the largest. We calculated the flow of the lowest 15 eigenvalues as a function of $m$ for two ensembles of 10 $L=8^3\times 16$ mean-field improved Symanzik configurations, one fine ensemble at $\beta = 4.80$ and one coarse ensemble at $\beta = 4.38$. The following flow graphs allow us to see the $m$ value for the biggest gap, and also allow us to compare the different actions. As we are interested in the magnitude of the low-lying values rather than their sign, we plot $|\lambda|$ vs $m$. The small physical extent of the fine lattices leads to qualitatively different behaviour in the spectral flow from the coarse lattice.

\begin{figure}[!tb]
\includegraphics[height=0.48\textwidth, width=0.28\textheight, angle=90 ]{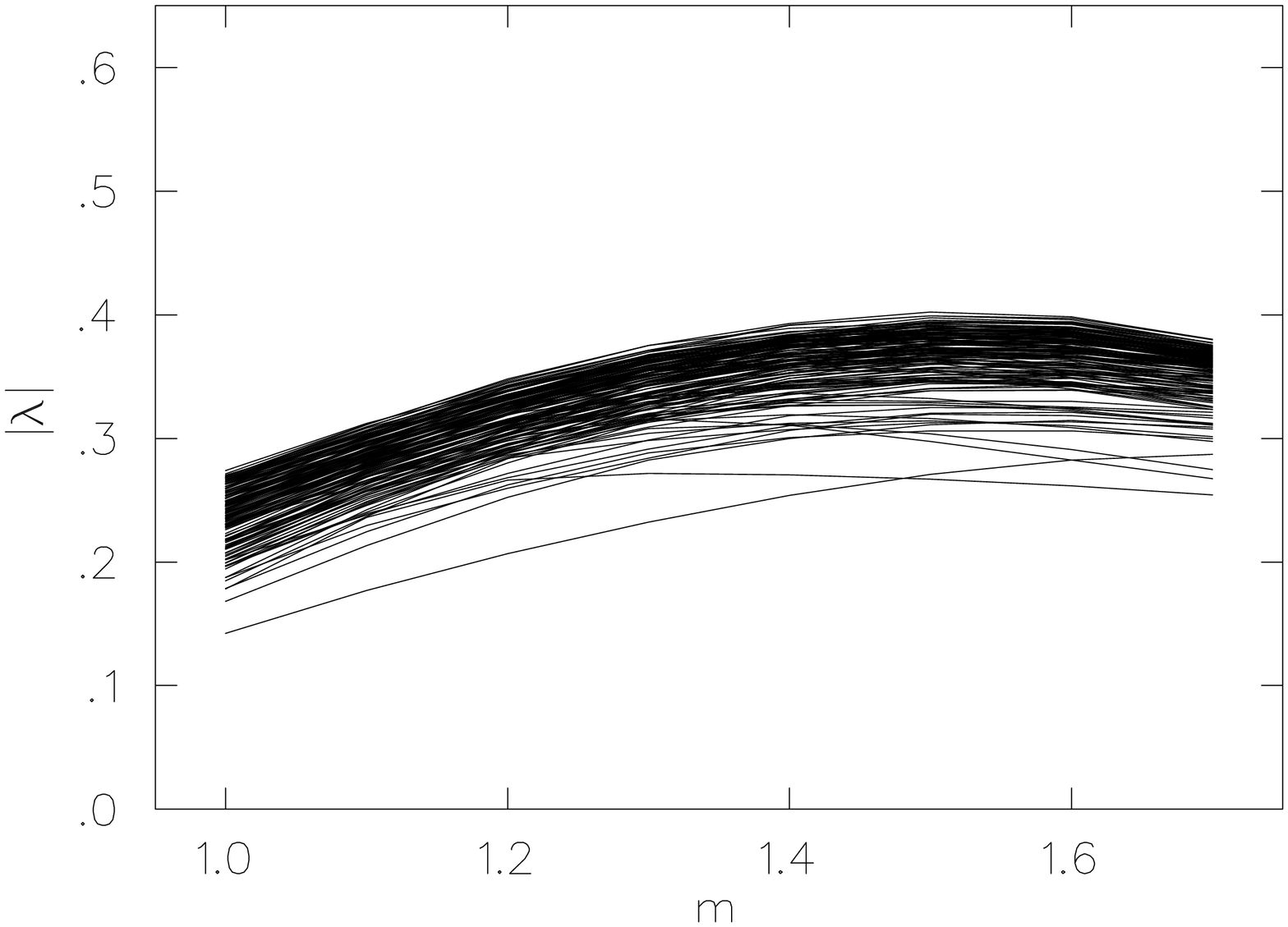}
\includegraphics[height=0.48\textwidth, width=0.28\textheight, angle=90 ]{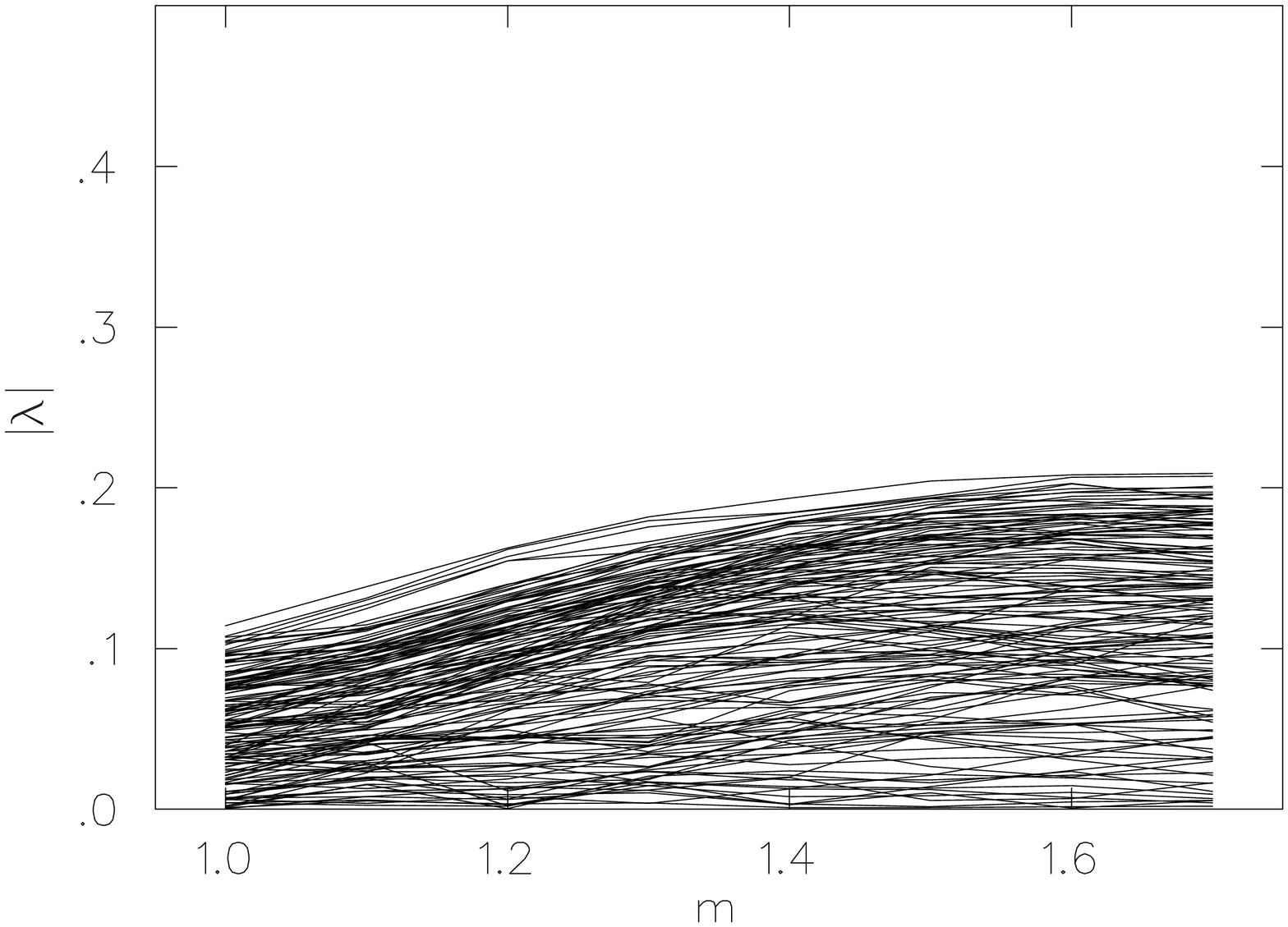}
\caption{\label{fig:wilson} Spectral flow of the Wilson action at $\beta=4.80$ (top) and 4.38 (bottom).}
\end{figure}

We begin by examining the flow of the Wilson action, in Figure \ref{fig:wilson}. We see on the fine lattices that there is a reasonable gap away from zero and a handful of isolated low-lying eigenmodes before the spectra becomes dense. On the coarse lattices the Wilson spectra is very poor, with a high density of very small eigenmodes. 

\begin{figure}[!tb]
\includegraphics[height=0.48\textwidth, width=0.28\textheight, angle=90 ]{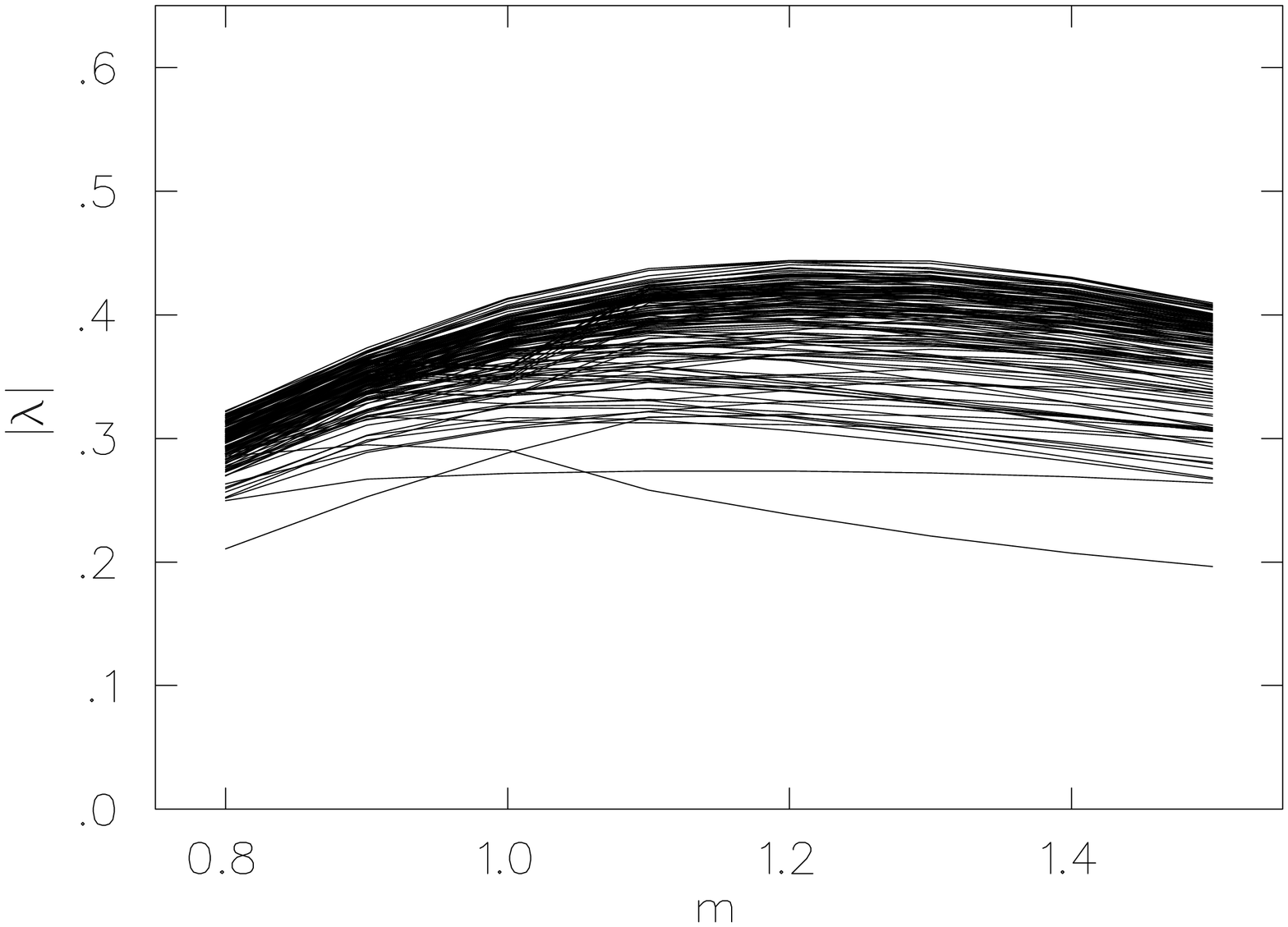}
\includegraphics[height=0.48\textwidth, width=0.28\textheight, angle=90 ]{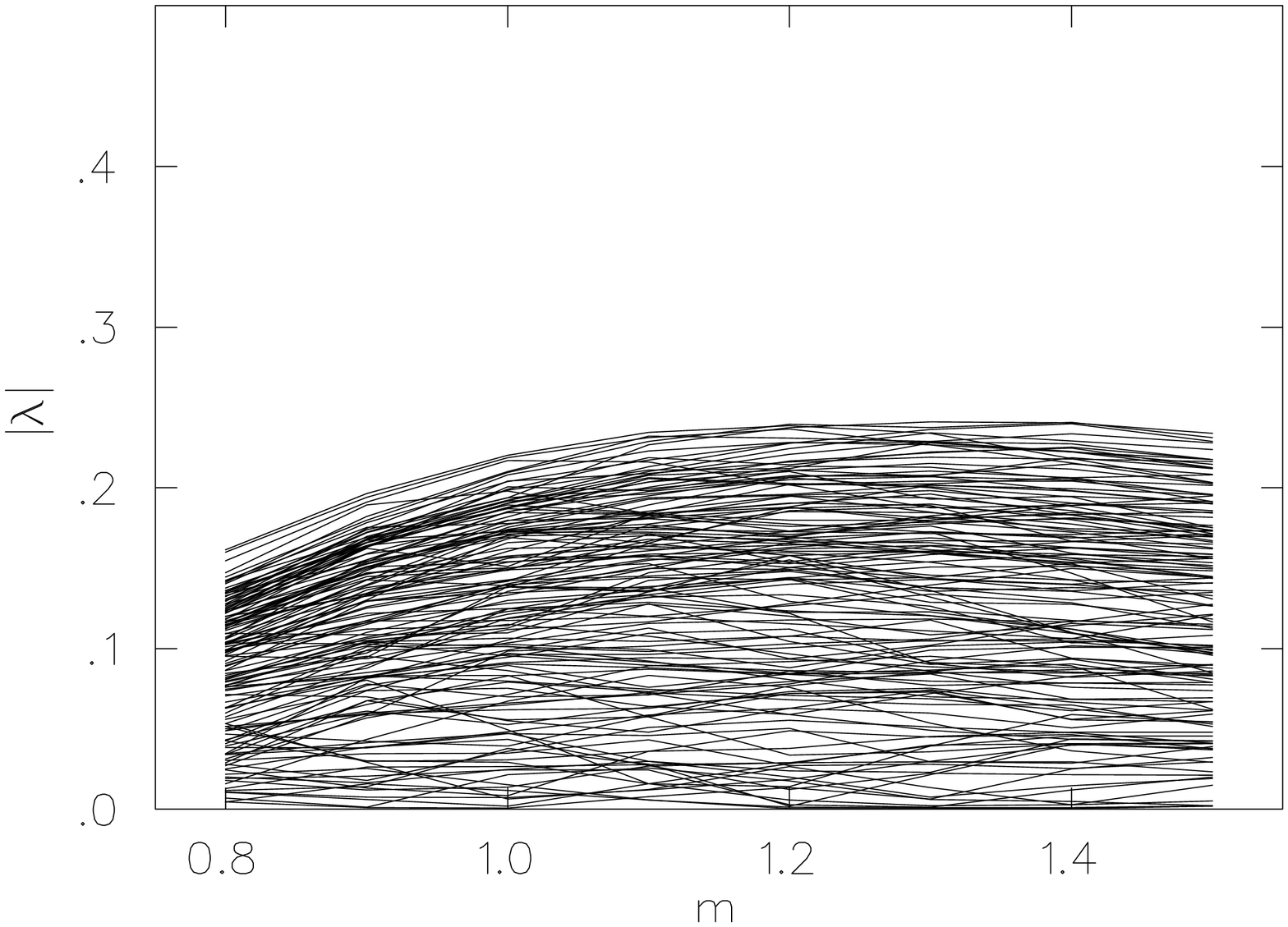}
\caption{\label{fig:clover} Spectral flow of the clover action at $\beta=4.80$ (top) and 4.38 (bottom).}
\end{figure}

The addition of the clover term (at $c_{\rm sw}=1$) provides some improvement (Figure \ref{fig:clover}), shifting the flow upwards on the fine lattices and moving the peak gap to $m=1$ as expected. The clover action still performs poorly on the coarse lattices however, as the presence of many small eigenmodes persists, although their density is clearly reduced. 

\begin{figure}[!tb]
\includegraphics[height=0.48\textwidth, width=0.28\textheight, angle=90 ]{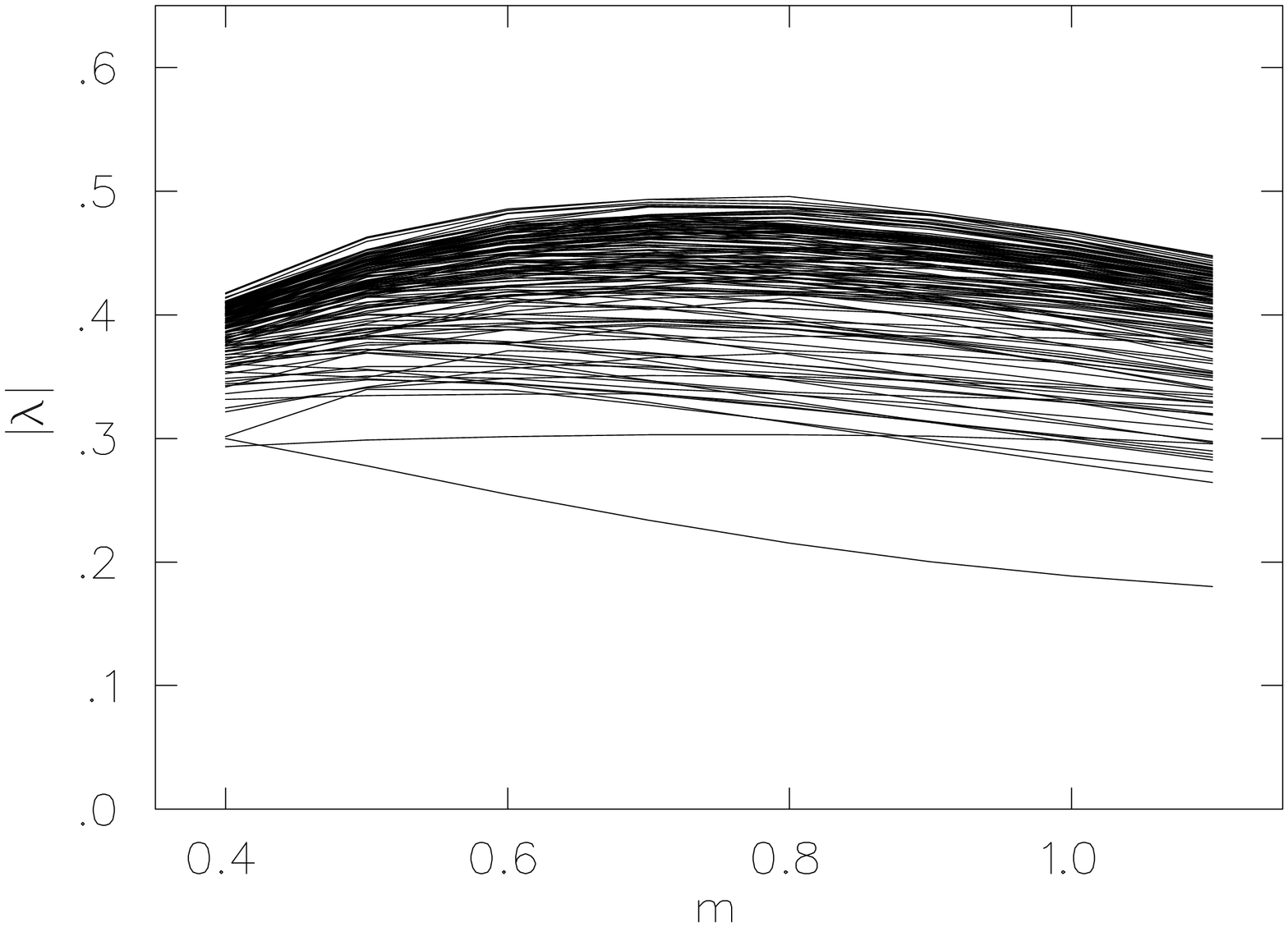}
\includegraphics[height=0.48\textwidth, width=0.28\textheight, angle=90 ]{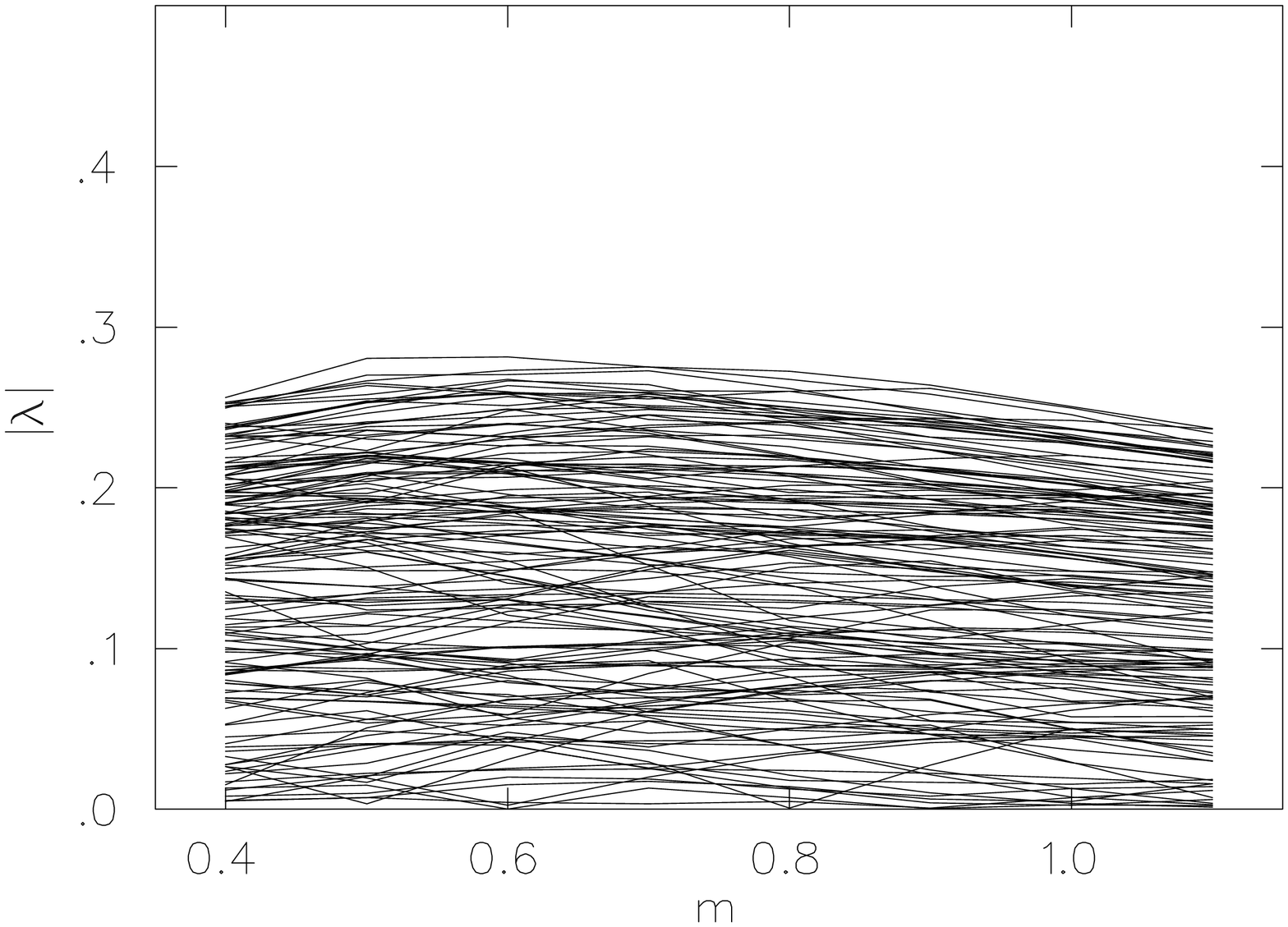}
\caption{\label{fig:mficlover} Spectral flow of the MFI clover action at $\beta=4.80$ (top) and 4.38 (bottom).}
\end{figure}

Mean field improvement assists the basic clover action somewhat (Figure \ref{fig:mficlover}), spreading the spectrum upwards, although the lowest modes are not raised significantly. The mass value at which the peak gap occurs has moved significantly from $m=1$ to $m=0.6$. As mentioned earlier, essentially all MFI does in this case is to change the value of $c_{\rm sw}$ to $1.0/u^3_0$, pushing it towards its non-perturbative value.
   
\begin{figure}[!tb]
\includegraphics[height=0.48\textwidth, width=0.28\textheight, angle=90 ]{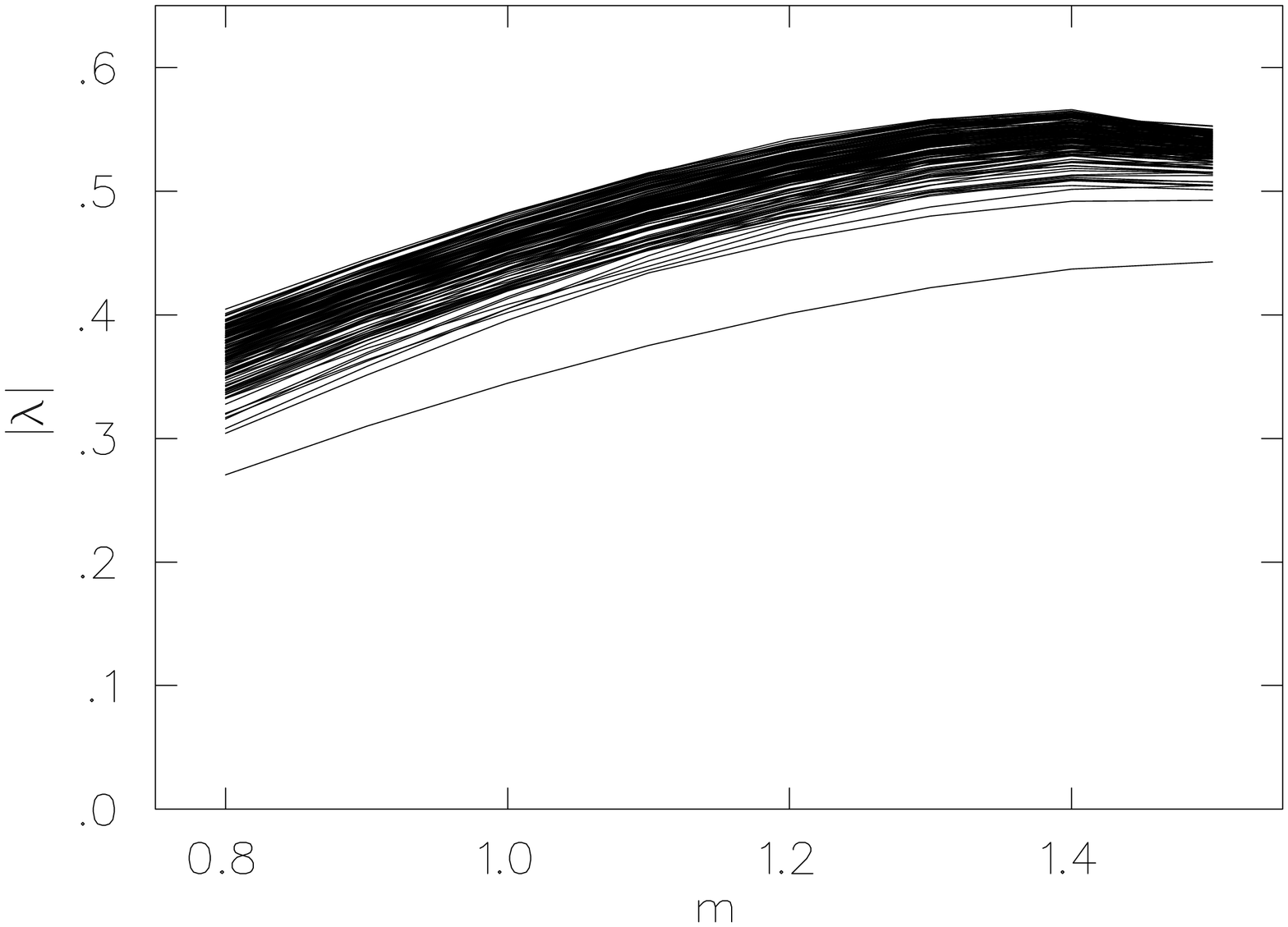}
\includegraphics[height=0.48\textwidth, width=0.28\textheight, angle=90 ]{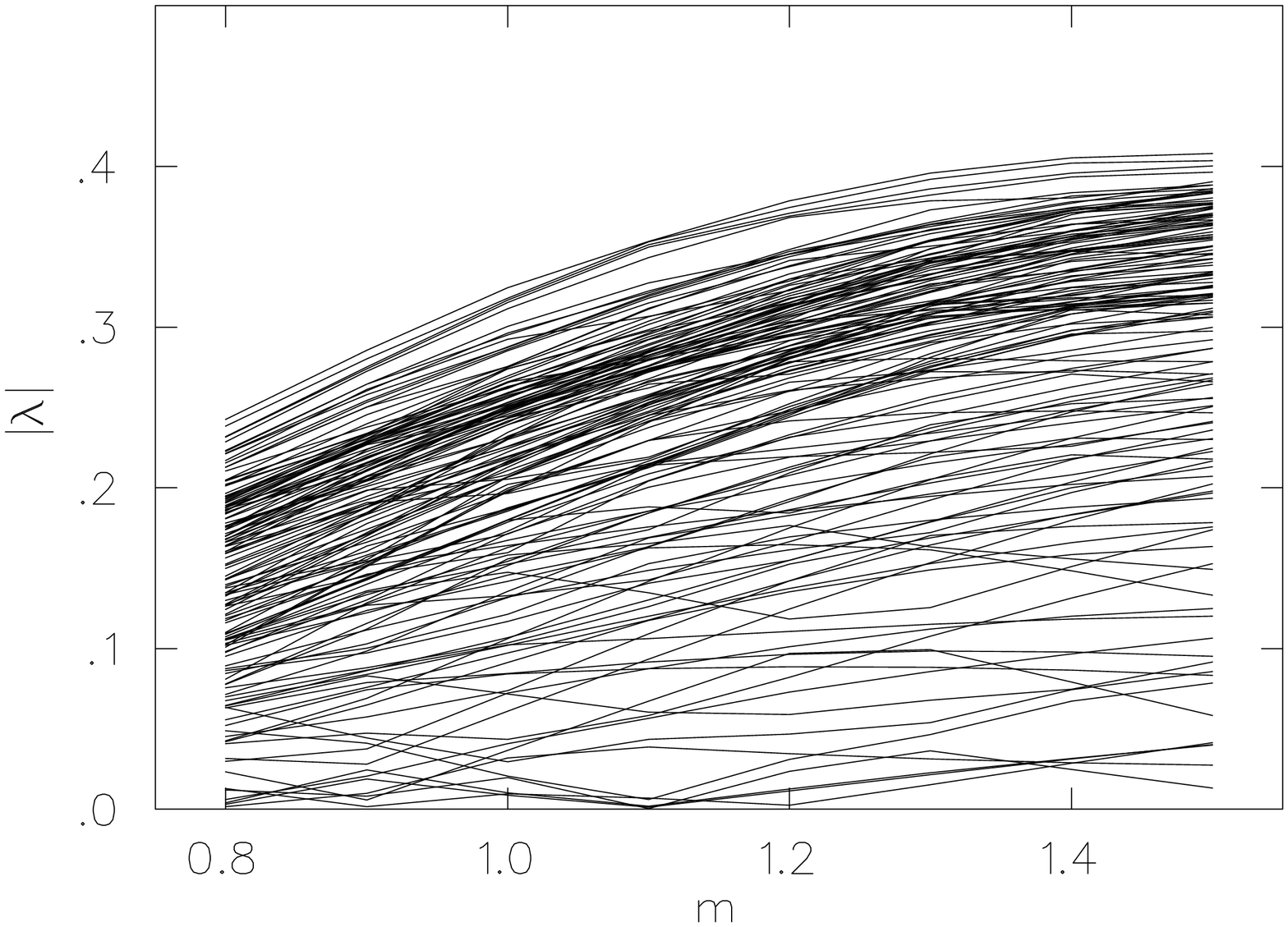}
\caption{\label{fig:fatwilson} Spectral flow of the fat Wilson action at $\beta=4.80$ (top) and 4.38 (bottom).}
\end{figure}

Modifying the Wilson action by smearing the irrelevant operators provides a considerable improvement (Figure \ref{fig:fatwilson}), enhancing the gap around zero on the fine lattices significantly. While there are still some small modes present on the coarse lattices, their density has been greatly reduced, and the spectral flow now has a clear division between the isolated low-lying modes and the modes where the spectral density becomes high. The latter are well separated from zero. Smearing was performed with $\alpha=0.7$ and $n_{\rm ape}=12$ smearing sweeps.

\begin{figure}[!tb]
\includegraphics[height=0.48\textwidth, width=0.28\textheight, angle=90 ]{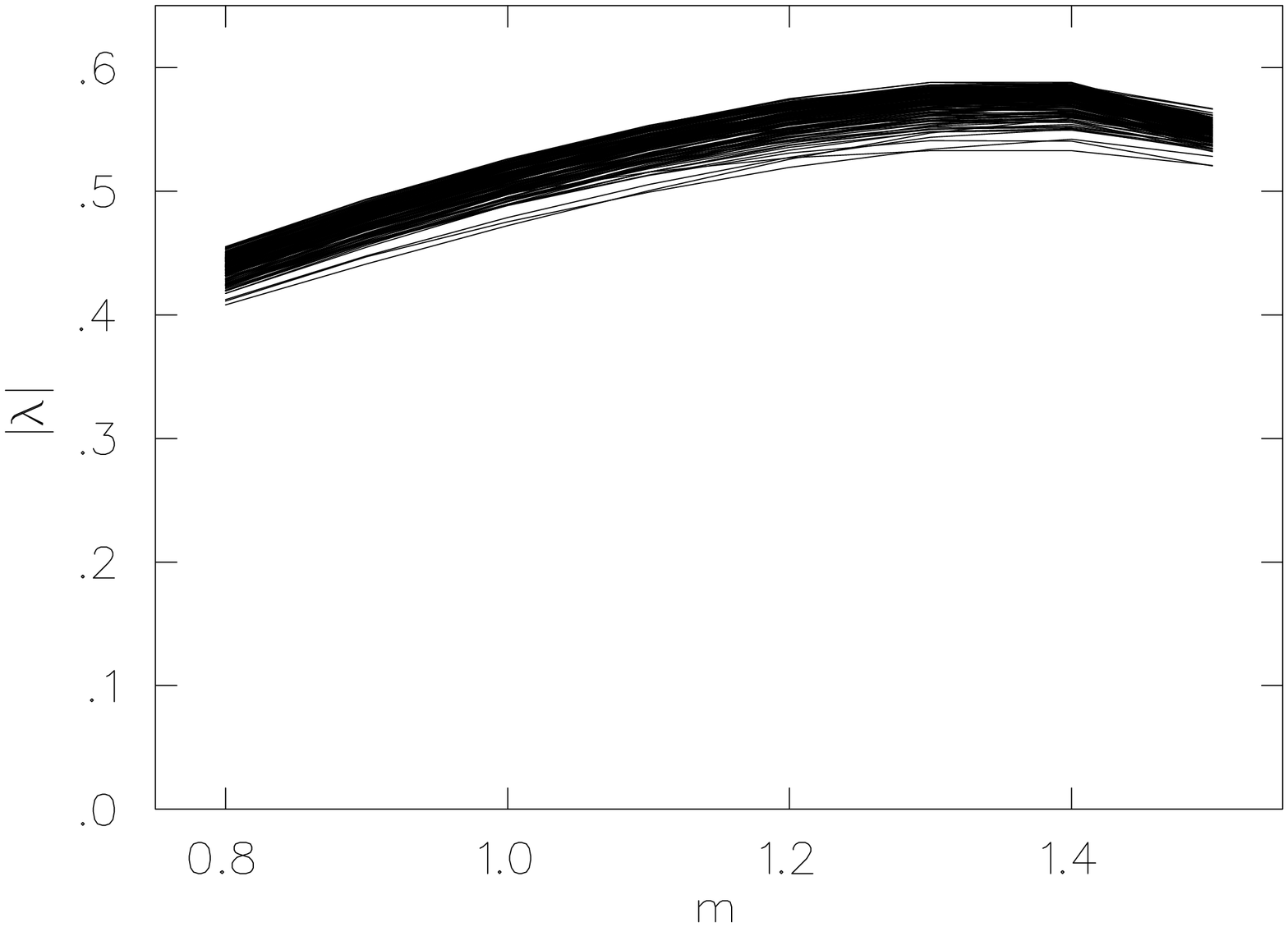}
\includegraphics[height=0.48\textwidth, width=0.28\textheight, angle=90 ]{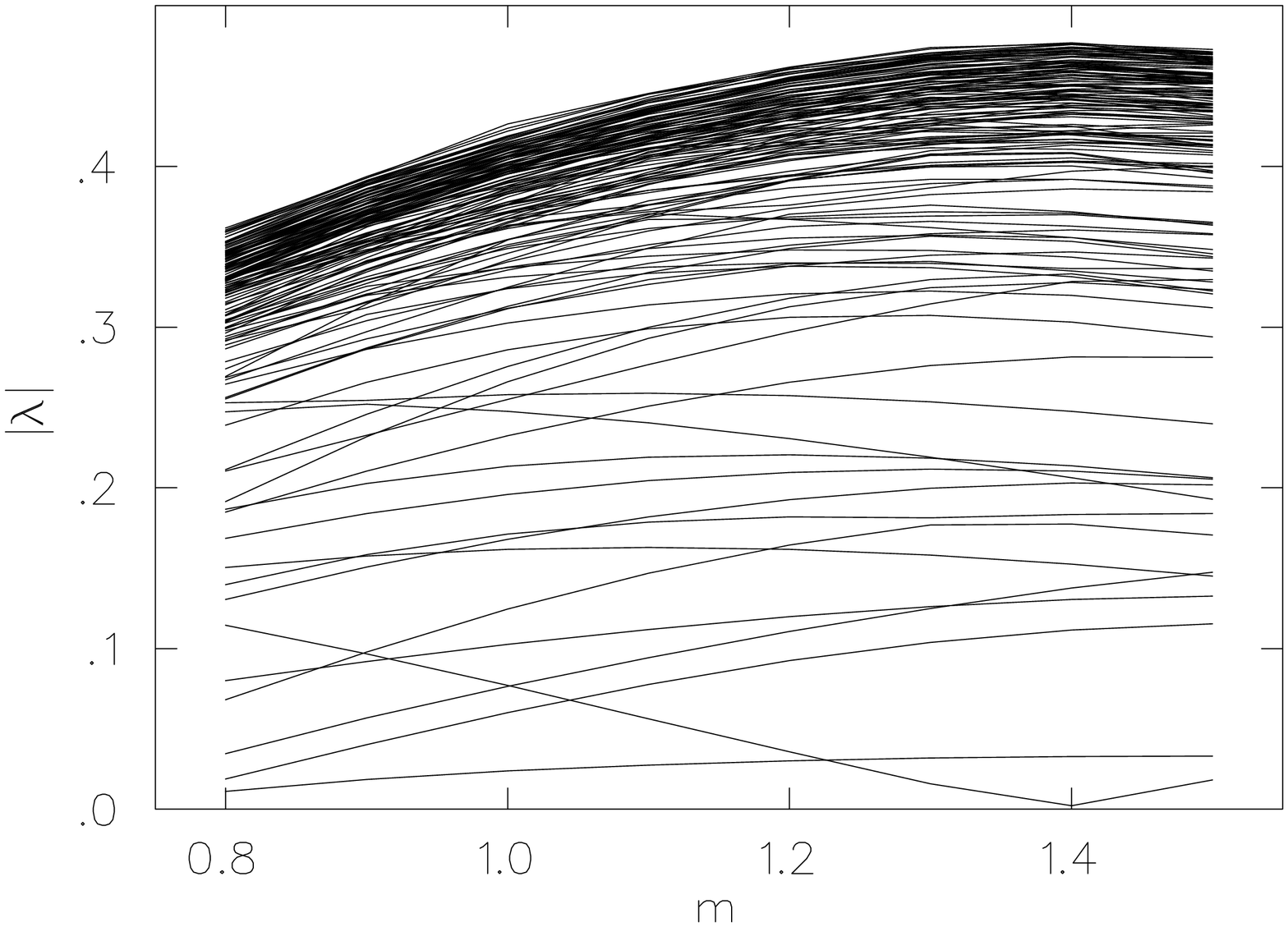}
\caption{\label{fig:fatclover} Spectral flow of the fat clover action at $\beta=4.80$ (top) and 4.38 (bottom).}
\end{figure}

The spectral flow of the fat clover action clearly demonstrates the superiority of clover-improved actions (Figure \ref{fig:fatclover}). The gap around zero is enhanced again over the fat Wilson action, and the number of isolated low-lying modes is reduced. The effect is very pronounced on the coarse lattice. 

\begin{figure}[!tb]
\includegraphics[height=0.48\textwidth, width=0.28\textheight, angle=90 ]{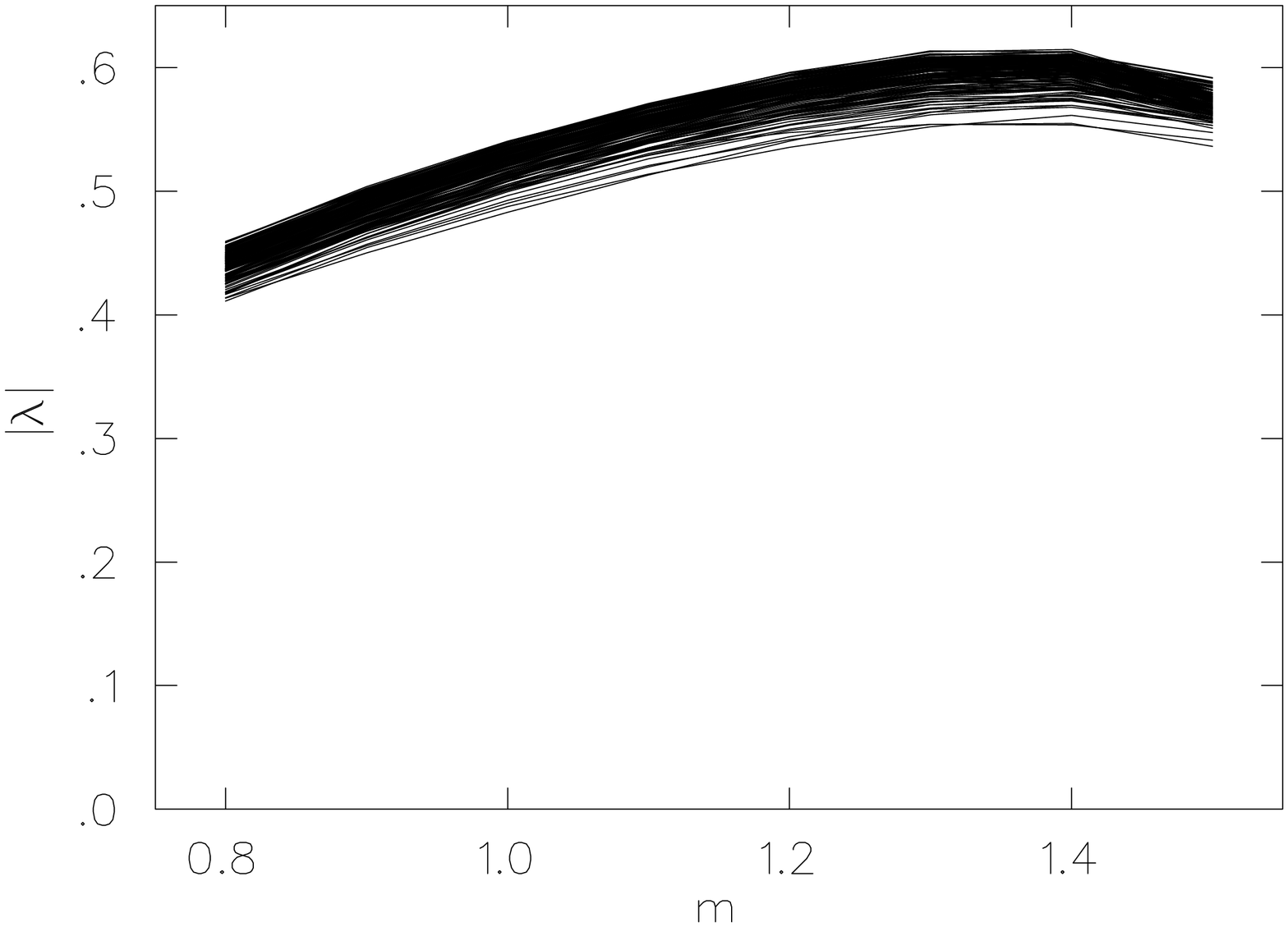}
\includegraphics[height=0.48\textwidth, width=0.28\textheight, angle=90 ]{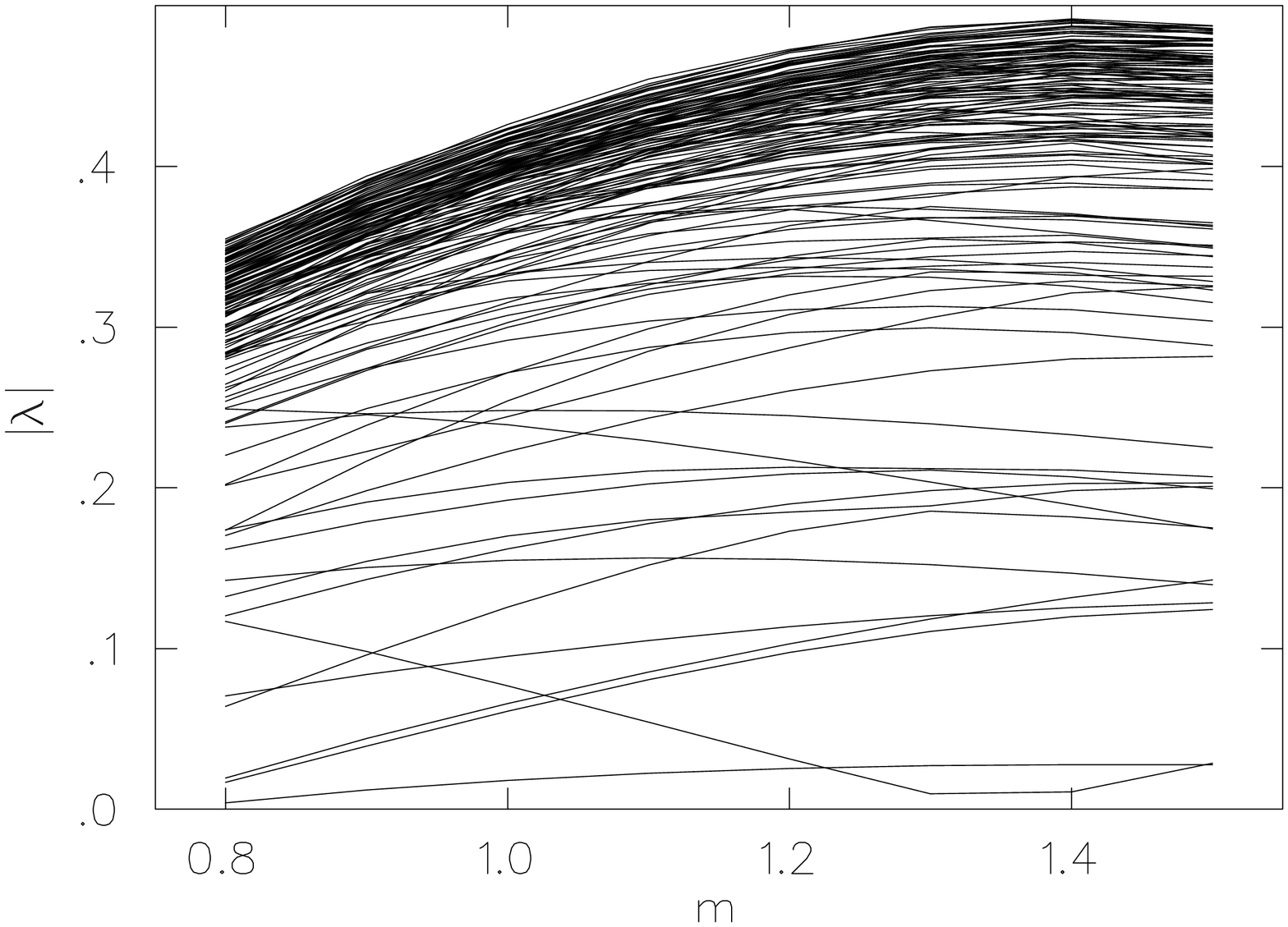}
\caption{\label{fig:flic} Spectral flow of the FLIC action at $\beta=4.80$ (top) and 4.38 (bottom).}
\end{figure}

As the fat links are already close to unity, the addition of mean field improvement only affects the fat clover flow slightly (Figure \ref{fig:flic}), raising the gap around zero a little and spreading the eigenvalues upwards slightly also. The low-lying density is again good in this case and far superior to that of the Wilson in Figure \ref{fig:wilson}. On the fine lattices we note that we have a peak gap at $m=1.4$ of $0.5$! On the coarse lattices the region where the eigenmodes start becoming dense occcurs at about $0.35$ at the same mass value, indicating that the FLIC action should be extremely well-conditioned on both ensembles, especially so on the coarse lattice if we project out the isolated low-lying modes. 

\begin{figure}[!tb]
\includegraphics[height=0.48\textwidth, width=0.28\textheight, angle=90 ]{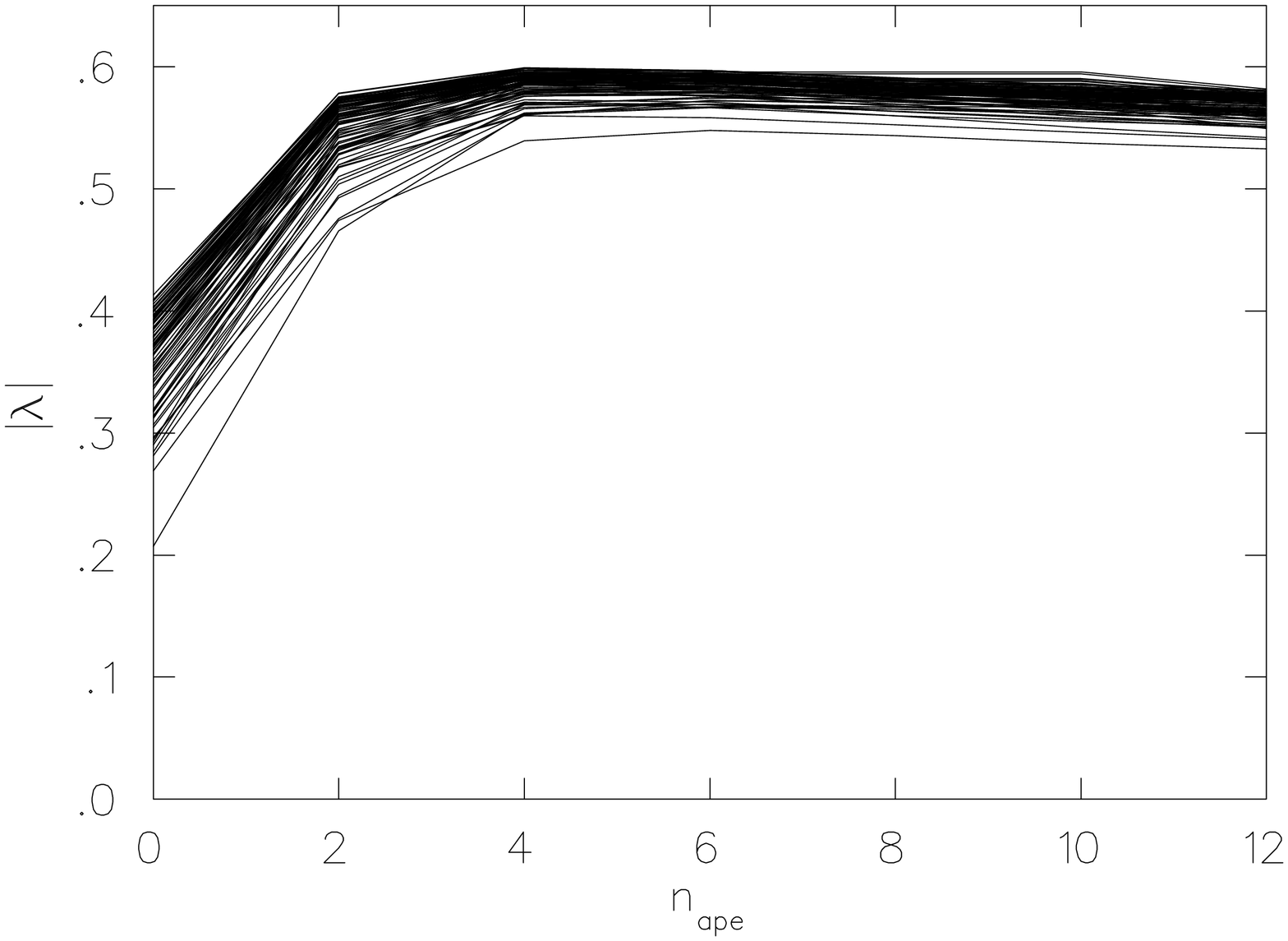}
\includegraphics[height=0.48\textwidth, width=0.28\textheight, angle=90 ]{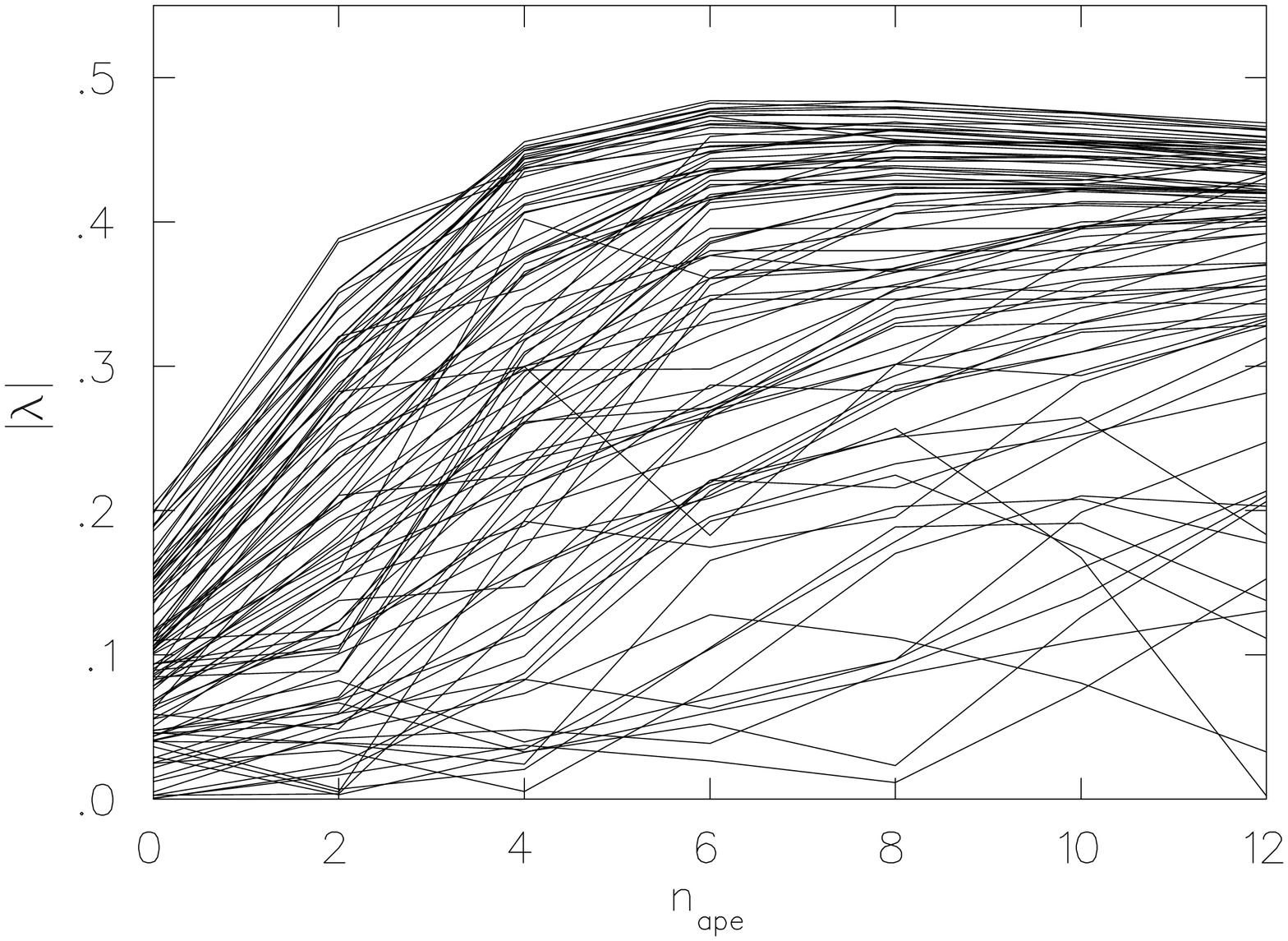}
\caption{\label{fig:smearfatclover} Dependence of the fat clover spectrum at $\beta=4.80$ (top) and 4.38 (bottom) on the number of APE smearing sweeps.}
\end{figure}

\begin{figure}[!tb]
\includegraphics[height=0.48\textwidth, width=0.28\textheight, angle=90 ]{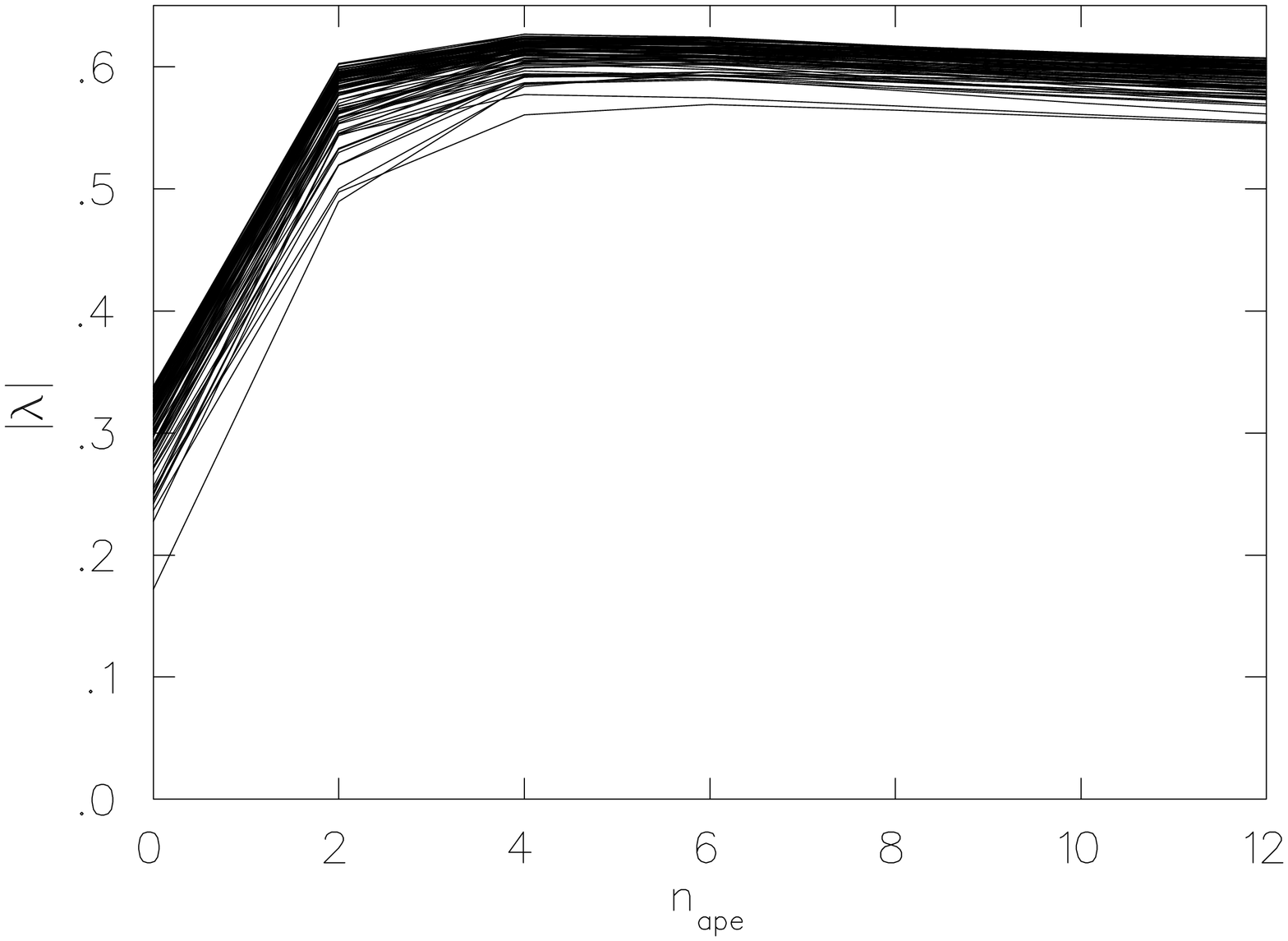}
\includegraphics[height=0.48\textwidth, width=0.28\textheight, angle=90 ]{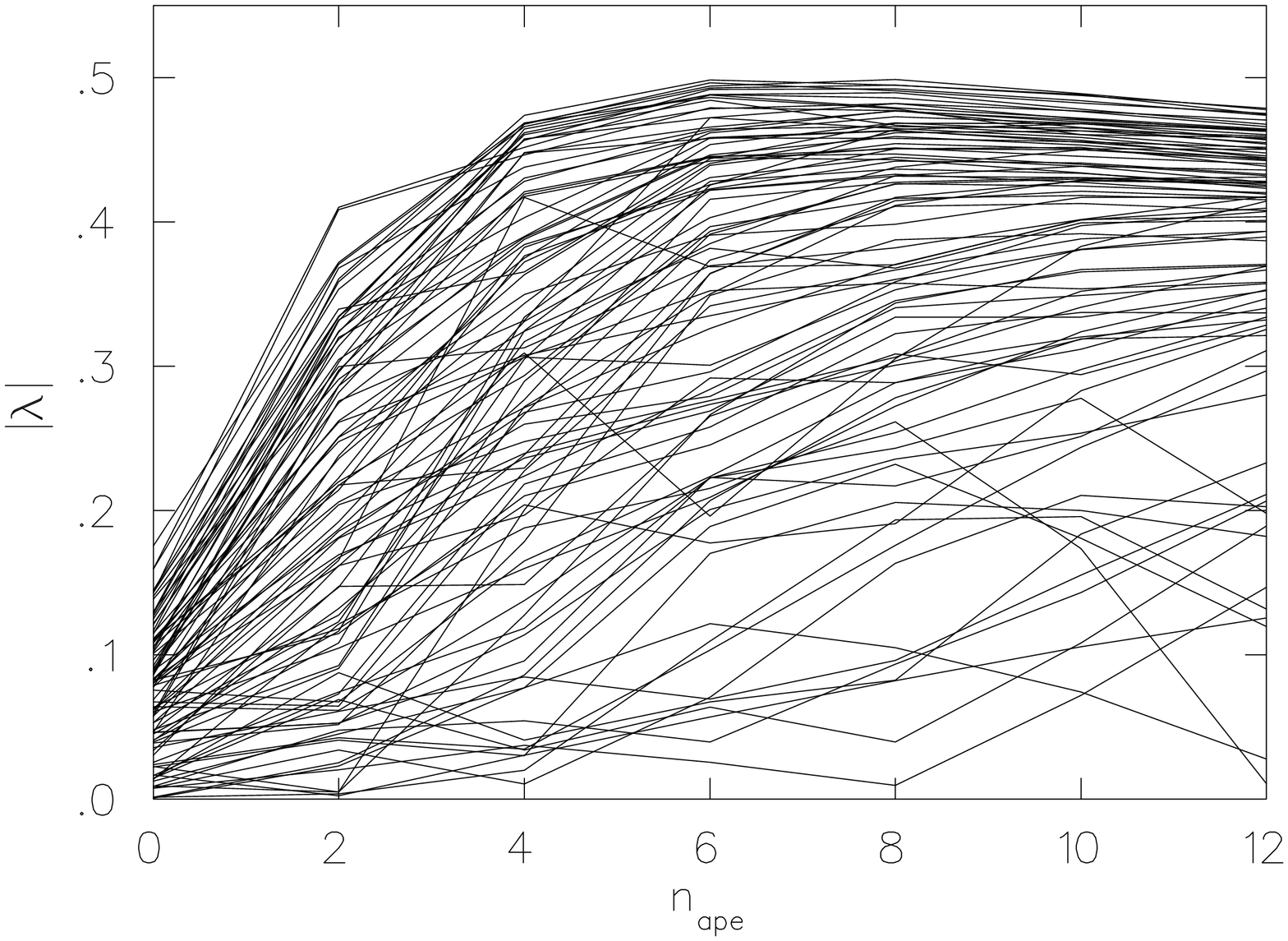}
\caption{\label{fig:smearflic} Dependence of the FLIC spectrum at $\beta=4.80$ (top) and 4.38 (bottom) on the number of APE smearing sweeps.}
\end{figure}

Additionally, we tested the dependence of the fat clover and FLIC actions upon the amount of smearing done (Figures \ref{fig:smearfatclover},\ref{fig:smearflic}). As stated in \cite{derek-apesmearing}, we only effectively need to vary the product $\alpha n_{\rm ape}$, so we fix $\alpha$ at 0.7 and vary $n_{\rm ape}$ between 0 and 12. We observe that in both the case of the fat clover and FLIC actions the initial 4-6 sweeps have a significant effect, but past 6 sweeps the effect is marginal, with the low lying density remaining constant and the eigenvalues being compressed very slightly downwards.
     
\section{Conclusion}

The addition of the clover-term at tree-level or with a mean field improved coefficient does not drastically improve the low-lying spectrum of the Wilson action at an optimal choice of $m$. However, APE smearing the irrelevant-terms in the Wilson action both with and without the clover term provides a marked improvement in the low-lying spectrum, reducing the number of isolated low-lying modes and shifting the spectrum away from zero. The combination of APE-smearing and the clover term at a mean-field improved coefficient provides the best overall low-lying spectra. Thus we conclude that the FLIC action is the best candidate for the overlap kernel out of the six actions tested. These results are built upon in \cite{kamleh-accelerated} where a comprehensive numerical comparsion is undertaken.

\bibliographystyle{h-elsevier.bst}

\bibliography{reference}





\end{document}

%% file: commands.tex
\usepackage{amsmath}
\usepackage{amssymb}
\usepackage{amsfonts}
\usepackage{amsthm}

\newtheorem{theorem}{Theorem}[section]
\newtheorem{corollary}[theorem]{Corollary}
\newtheorem{lemma}{Lemma}[section]
\newtheorem{definition}{Definition}[section]
\newtheorem{proposition}{Proposition}[section]
\newtheorem{example}{Example}[section]

\newcommand{\eqn}[1]{ \begin{equation} #1 \end{equation} }

%% file: cairnsproc.bbl
\begin{thebibliography}{1}

\bibitem{neuberger-practical}
H. Neuberger,
\newblock Phys. Rev. Lett. { 81} (1998) 4060, hep-lat/9806025.

\bibitem{edwards-practical}
R.G. Edwards, U.M. Heller and R. Narayanan,
\newblock Nucl. Phys. { B540} (1999) 457, hep-lat/9807017.

\bibitem{edwards-chiral}
R.G. Edwards, U.M. Heller and R. Narayanan,
\newblock (1999), hep-lat/9905028.

\bibitem{kamleh-accelerated}
W. Kamleh et~al.,
\newblock (2001), hep-lat/0112041.

\bibitem{zanotti-hadron}
J.M. Zanotti et~al.,
\newblock (2001), hep-lat/0110216, and these proceedings.

\bibitem{edwards-spectral}
R.G. Edwards, U.M. Heller and R. Narayanan,
\newblock Nucl. Phys. { B535} (1998) 403, hep-lat/9802016.

\bibitem{derek-apesmearing}
F.D.R. Bonnet et~al.,
\newblock Phys. Rev. D62 (2000) 094509, hep-lat/0001018.

\end{thebibliography}
